# Fold or die: experimental evolution in vitro

Sinéad Collins[1], Andrew Rambaut[1], Stephen J. Bridgett[2]

Data archiving: sequence data archived on datadryad.org

1. University of Edinburgh
Institute of Evolutionary Biology
Ashworth Laboratories, The King's Buildings
West Mains Road
Edinburgh EH9 3JT

2. University of Edinburgh
Institute of Evolutionary Biology, The Gene Pool Genomics Facility
Ashworth Laboratories, The King's Buildings
West Mains Road
Edinburgh EH9 3JT

Email: s.collins@ed.ac.uk

**Abstract**

We introduce a system for experimental evolution consisting of populations of short oligonucleotides (Oli populations) evolving in a modified quantitative Polymerase Chain Reaction (qPCR). It is tractable at the genetic, genomic, phenotypic and fitness levels. The Oli system uses DNA hairpins designed to form structures that self-prime under defined conditions. Selection acts on the phenotype of self-priming, after which differences in fitness are amplified and quantified using qPCR. We outline the methodological and bioinformatics tools for the Oli system here, and demonstrate that it can be used as a conventional experimental evolution model system by test-driving it in an experiment investigating adaptive evolution under different rates of environmental change.

A central goal of evolutionary biology is to explain adaptation seamlessly from gene to ecosystem – to identify genetic changes within a population, quantify the action of natural selection on genetic variation, link this genetic variation to phenotypic variation and to variation in fitness, and finally, to relate organismal changes to adaptation to some aspect of the environment. To do this comprehensively in a real population in a real ecosystem requires persistence and luck. The handful of successes study traits with a simple genetic basis underlying a phenotype with a clear and easily measured adaptive value, or involve research lifetimes devoted to the careful study of a single natural system (Abzhanov et al. 2006; Schluter et al. 2010). One alternative to such luck and devotion is to bring evolution into the lab, where biologists can carefully control environmental conditions and use genetically-tractable model organisms to study traits that have clear links to fitness particular environmental drivers. This approach has been successful with systems such as *Pseudomonas*, where adaptive radiation can be reliably induced under laboratory conditions (Spiers et al. 2002; McDonald et al. 2009). However, we are still unable to predict or interpret most of the genetic variation that occurs during lab selection experiments, even in our best-studied model organisms evolved in monocultures in simple environments, and many of our insights stem from case studies of exceptions rather than systematic surveys (Barrick et al. 2009). Although high-throughput sequencing can give us a comprehensive picture of genetic changes, it cannot help us link these genetic changes to changes in phenotype or to particular environmental drivers. In short, published studies have provided us with a sample of what can happen in adapting populations, but have left open the question of what, on average, does happen most of the time.

Much of our inability to explain adaptation from gene to ecosystem stems from our lack of data on the distributions of effects that contribute to fitness gain. Concretely, it would be useful to know the distributions of fitness effects of beneficial mutations, of fixed mutations, and of epistatic interactions between fixed mutations. For example, knowing the distribution of epistatic interactions between fixed mutations would allow us to quantify epistatic constraints on evolutionary trajectories and formalize how epistasis and pleiotropy contribute to the repeatability of adaptive outcomes.

Here, we present a new *in-vitro* system based on a modified quantitative Polymerase Chain Reaction (qPCR) for carrying out laboratory selection experiments to build empirical genotype-phenotype-fitness maps and to measure the distributions of fitness effects of mutations and epistatic interactions for adapting populations of DNA hairpins. This system is a halfway-point between computer simulations, which are limited by our understanding of how biology works, and simple viruses or virions, which, though simple relative to cellular organisms, are often still too complex to be completely tractable (Sanjuán 2010; Peris et al. 2010) . Our system is conceptually similar to RNA-based *in-vitro* systems such as the Qβ system (Orgel 1079; Joyce 2007), where short molecules evolve under carefully defined chemical conditions and can be studied at the level of sequence change or phenotypic change.

While RNA-based or aptamer systems have been used to demonstrate that natural selection operates in vitro (Orgel 1079), shed light on origin of life chemistry and the evolution of the specific biology of short nucleotides (Ellington 2009), they have not been widely adopted as a model system for experimental evolution studying fundamental questions on the distributions of fitness effects of mutations, epistatic interactions, or determinants of lineage success or extinction in the same way or to the same extent that *in-silico,* viral, and microbial systems have, and are now used primarily by and for chemistry researchers. We hope that by introducing a PCR-based, easy-to-use system that depends on skills and equipment that most experimental evolutionary biologists already possess, *in-vitro* evolution will become more widely used. Indeed, our system could be used alongside one like Qβ to gain power and generality, in much the same way *Pseudomonas* is used alongside *E.coli*.

The *in-vitro* system presented here, like other nucleotide-based systems for *in-vitro* evolution, has shortcomings, one of which is obviously that it is far simpler than any real organism. However, it is intended to fill in gaps in our current arsenal of experimental and digital model systems used to ask general questions using experimental evolution, not to replace any of them. For this proof-of-principle study, we use this novel *in-vitro* model system to explore determinants of lineage persistence in changing environments, and outline other questions that could be answered from the same experiment.

Figure 1

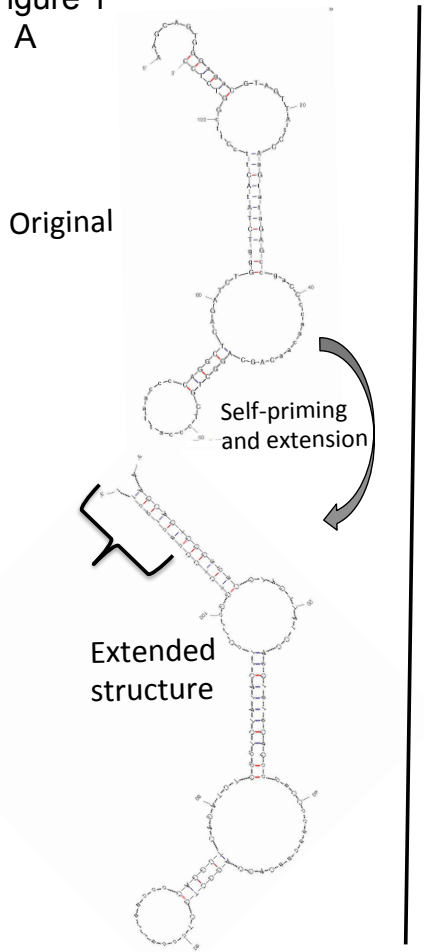
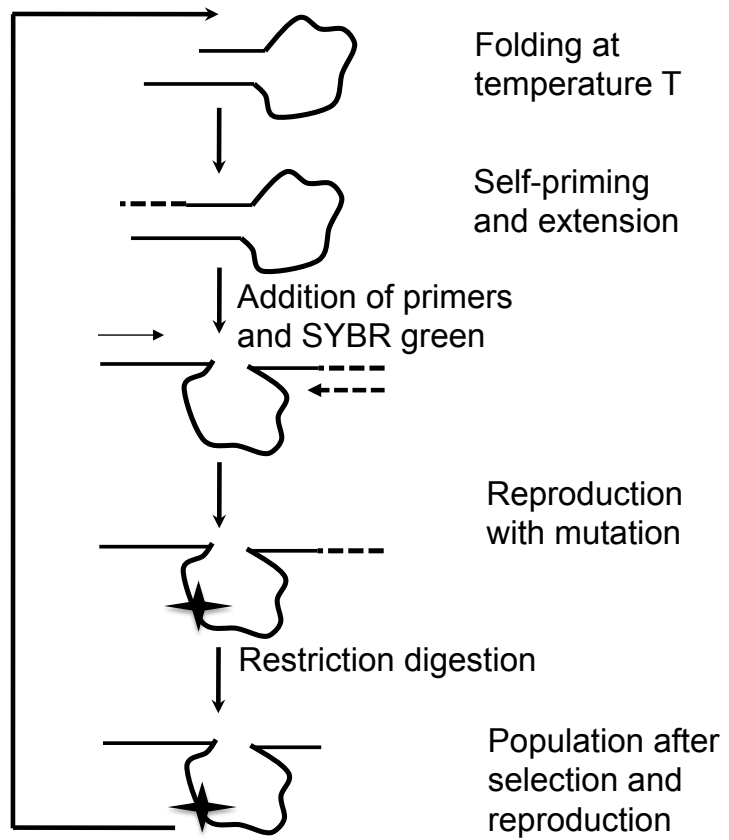

**Figure 1** Schematic of the founding Oli molecule and "fold or die" selection. A: Self-priming and extension in the founding Oli molecule. The most likely structure of Oli for folding and self-priming at the beginning of the experiment is shown. See Methods for details of structure prediction. B: A single cycle of "fold or die" selection. Oli molecules are challenged to fold at a given temperature into a structure where they can self-prime and extend. Populations of extended molecules grow exponentially in the qPCR step, while populations of unextended molecules grow linearly. Following qPCR, the extended ends of Oli molecules are cleaved. These cleaved populations are diluted and a subsample is used for the next cycle of "fold or die" selection.

**The model system**
The "organism" is called Oli, which is short for "short oligonucleotide". Oli is a single-stranded DNA hairpin designed to fold stably at 45°C in its ancestral state into the predicted hairpin structure shown in Figure 1A, where it is the only likely structure. This predicted hairpin structure becomes less stable at higher temperatures until it melts completely above 55°C, so that it is unlikely to form a hairpin where the two ends manage to pair over the section where they are complimentary. The hairpin is still predicted to form with low likelihood with a negative free energy at 55°C. The upper temperature limit for self-priming was determined empirically by quantitative PCR, and is consistent with biomolecular folding predictions. Details of Oli design and structure can be found in the Methods. Since Oli has a genotype (sequence) and a phenotype (predicted most likely structure) that is sensitive to environmental conditions, such as ion concentrations and temperature, Oli populations can evolve if they can be copied with a mutation rate. Oli does not reproduce by being alive, but is instead copied by being supplied with the building blocks for new individuals (deoxynucleotides) and the means to use them (Taq polymerase) under defined chemical and physical conditions that can be easily monitored (qPCR reactions). Mutagenic Taq with a defined mutation rate and spectrum can be used to control the rate of evolution of Oli populations.

Oli populations can be used for laboratory selection experiments as outlined in Figure 1B. Here, the trait under selection is the ability to form a structure that pairs the complimentary sections of the 5' and 3' ends of Oli at increasing temperatures. A single round of selection and reproduction with mutation consists of the following five steps: **1.** In a solution supplied with deoxynucleotides and Taq, initially dilute Oli populations are given a short time to form any structure that allows them to self-prime at a given temperature. Those that are able to self-prime can extend their 3' end (indicated by bracket in figure 1A and a dashed section in 1B). Because populations are dilute at this step it is more effective to self-prime than to use another Oli molecule as a primer, though using other molecules as primers probably happens at low frequencies. **2.** Primers are added to the solution. Molecules that have successfully self-primed and extended in step 1 are able to bind both primers, while molecules that have failed can only bind one primer. **3.** Oli populations expand in a PCR reaction. Molecules able to bind both primers found lineages that increase exponentially, while molecules that only bind one primer found lineages that increase linearly, giving a selective advantage to molecules that self-primed in step 1. This results in "fold or die" selection on phenotype. In this step, primer binding temperatures are kept constant, and a the formation of a secondary structure allowing self-priming is no longer needed. Overall population growth rates can be monitored in real time using SYBR green. Populations are primer limited. Details of the PCR reagents and program are given in the Methods. **4.** The extended (3') ends of the Oli molecules are cleaved with two restriction enzymes. Restriction sites are designed so that escaping cleavage results in being unable to bind one primer, which selects against that strategy. **5.** Approximately $10^5$ individuals (1μl) from the diluted and digested population are used for the next round, where steps 1-5 are repeated. In principle, it is also possible to use media composition as a selective environment, but it is logistically simpler to manipulate reaction temperatures.

At the end of an experiment, populations from some or all of the time points can be sequenced using 454 sequencing. Oli is short enough so that no assemblies are required, and sequencing runs can be of near-entire populations, rather than small samples of populations, allowing population composition to be calculated directly from sequence data taken from a large proportion of the population rather than estimated from a small subsample. This allows rare or transient genotypes to be detected. Fitness for entire populations is measured during the selection experiment as qPCR output, and fitness for individual lineages/genotypes can be measured by synthesizing the derived Oli sequence of interest and subjecting them to steps 1-3 above. Oli is short enough that it may be ordered and synthesized as a standard oligonucleotide in most cases. When genome expansion occurs, Oli must be ordered as a synthetic gene.

This paper presents a "proof of principle" experiment to demonstrate how the system may be used. Populations of Oli were selected at increasing temperatures for self-priming, where the temperature was raised either suddenly or gradually. This experiment is analogous to previous studies that investigate evolutionary responses to different rates of environmental change in digital (Collins et al. 2007) and cellular (Collins and de Meaux 2009) model systems.

**Methods**

*Oli design and structure determination.* The unextended Oli molecule is a 107 bp oligonucleotide with the sequence 5'-AAGCAGTG**GGagaCCG**AGTTAtCCAaGtataGAGccgaCCccaacaaCAGCAGGCTGCTcccattaacccCAGGCTCAGATCTGggTCTAtACttcCtt**cGGTCTCC-**3'. Oli is based on a published DNA hairpin sequence from HIV-1 (Driscoll and Hughes 2000). Upper case letters indicate positions where Oli has the same sequence as the original hairpin, and lower case letters indicate positions where the Oli sequence differs from the original hairpin. To form a structure capable of self-priming, Oli molecules must pair the 3' end indicated in bold with the complimentary tract near the 5' end indicated in bold. Extension then produces a new 3' end … cCtt**cGGTCTC\*C\***CACTGCTT-3', capable of binding a primer supplied later in the protocol, and containing recognition sites for the two restriction enzymes BsaI and HpyAV (New England Biolabs) that cleave at the sites indicated by the asterisks. The recognition sequences of BsaI and HpyAV are GGTCTCN and CCTTC($N_6$) , respectively. To escape restriction digestion, 3 or more independent point mutations are needed. We verified that qPCR primers could not reliably bind and produce product with 2 or more mismatches under the stringent conditions used in our experiment. This selects against restriction digestion escape during the experiment. This is further confirmed by the failure of genotypes showing mutations that would allow "digestion escape" to show up in the sequenced populations, even at later timepoints. qPCR primer sequences are: 5'-aag cag tgg gag acc gaa gg-3' (Tm = 62ºC under reaction conditions used) and 5'-aag cag tgg gag acg tag tta tcc a-3' (Tm = 63ºC under reaction conditions used).

The most likely structure for the ancestral Oli was determined using mfold (http://mfold.rna.albany.edu/) with the following settings: single stranded DNA, 200 foldings maximum, temperature = 70ºC. Sequences as well as their reverse compliments were folded. Structures were not empirically verified, but the decreased ability of the original Oli molecules to self-prime and extend at temperatures above 55ºC was verified by sequencing the end products of PCR reactions that were carried out at increasing annealing temperatures for the self-priming step, which is the key feature needed for this system. The upper temperature limit for self-priming was also confirmed functionally using qPCR. PCR reactions were set up with a known number of starting molecules and subjected to a single round of selection (steps up to but not including restriction digestion) at self-priming temperatures from 45°C to 65°C followed by the addition of primers and SYBR green, then 35 cycles of standard qPCR. The number of starting molecules was then calculated from the qPCR data. When nearly all of the molecules are able to self-prime, the qPCR estimate of initial number of molecules matches the known initial number of molecules, since all molecules will be able to recruit both primers in all but the first cycle of the qPCR program. When few of the molecules self-prime, the starting population size estimated by qPCR is lower than the known starting population size of Oli. An extended Oli was synthesized and used as a positive control for the population growth rate expected if all of the molecules were able to self-prime.

*Selection experiment.* All populations were started from a single sample of Oli ordered from eurofins mwg|operon, diluted to a known concentration. The initial population was sequenced using conventional Sanger sequencing. Sequencing reads did not show polymorphisms, so initial diversity was presumed to be low relative to the diversity generated by the error-prone Taq during the experiment. All starting populations consisted of 1μl of 0.05pmol stock (about 30 000 molecules). Fifty-six independent replicate populations were used for each of the following four selection regimes: 1. control (no change in self-priming temperature); 2. sudden environmental change (a single increase in self-priming temperature of 15°C); 3. rapid environmental change (five sequential increases in self-priming temperature of 3°C each); 4. slow environmental change (ten sequential increases in self-priming temperature of 1.5°C each). In the three treatments where temperature increased, the total increase in temperature over the entire selection experiment is the same (15°C, from 55°C to 70°C); only the rate differs. See Supplementary Table S1 for a timeline of self-priming temperatures.

*PCR.* A quantitative mutagenic PCR kit was assembled from the components of the Brilliant II SYBR Green QRT-PCR kit, replacing the standard Taq with Mutazyme II from the GeneMorph II Random Mutagenesis Kit (Stratagene, California). Betaine was added to the reactions so that hairpins would melt. Information on the mutation rate and spectrum of Mutazyme II can be found at (http://www.genomics.agilent.com/). The PCR protocol consisted of one cycle of self-priming and extension, followed by the addition of primers and SYBR green to the reactions, followed by 40 cycles of quantitative PCR not requiring self-priming. The self-priming reaction used 1μl of (restriction digested) Oli, 7.08μl $H_2O$, 1.5μl 5M betaine, and the following volumes of reagents supplied in the Brilliant II SYBR Green QRT-PCR kit: 1.5μl buffer, 0.0932μl $MgCl_2$, 0.6μl dNTPs, 2.4μl glycerol, 0.45μl DMSO, 0.375μl Mutazyme II, for a total volume of 15μl per reaction. Program: melting at 98°C for 3min, hairpin folding temperature for 10sec, extension at 72°C for 2min. Following the self-priming and extension step, the reactions were cooled to 4°C in the PCR machine, moved to ice, and a cocktail made up of the following reagents was added to each well: 0.5μl 5M betaine, 0.5μl buffer, 0.0313μl $MgCl_2$, 0.2μl dNTPs, $5\times10^6$ molecules of each primer, 0.8μl glycerol, 0.15μl of (1:1500 diluted) SYBR green, 0.2μl Mutazyme II, bringing the total reaction volume to 20μl. Plates were spun briefly and returned to the machine. Program: initial melting at 98°C for 1 min followed by 40 cycles of: melting at 98°C for 1min, primer annealing at 67°C for 1 min extension at 72°C for 2 minutes. A melting curve was done for each reaction following this. At an annealing temperature of 67°C, a single mismatch is enough to impair primer binding. Except for betaine and Mutazyme II, all reagents were used in the concentrations supplied in the Brilliant II SYBR Green QRT-PCR kit. Providing each reaction with $5 \times 10^6$ molecules of each primer results in transfer population sizes of Oli of about $10^5$ individuals.

For qPCR reactions, water was used as a negative control. Standards for qPCR quantification were six different known concentrations of pre-extended Oli molecules that did not require self-priming, so that every member of the population could recruit both PCR primers. In fitness measurements, all growth rates and population sizes are calculated relative to standards.

*Restriction digestion.* Following PCR, populations were digested for 1 hour with BsaI, followed by digestion for 1 hour with HpyAV, followed by heat inactivation at 65°C for 25 minutes. 1µl of this diluted and digested Oli population (on the order of $10^5$ individuals) was used for the next round of PCR. Since populations are upstream (5'end) primer-limited, total population sizes are smaller than expected based on the number of PCR cycles that they have gone through. Maximum population size can be controlled by the amount of primer supplied and remains roughly constant over the selection experiment.

*Fitness measurements and population size determination.* The initial total population size is calculated fluorometrically using the Quant-iT DNA Assay kit, high sensitivity (Invitrogen) and black FluoroNunc 96-well flat-bottomed polystyrene microplates (Fisher). The standard Quant-iT protocol (20uL sample added to 180uL working solution) produced weak signals, possibly because folded and/or short molecules take up the fluorophore unevenly. We modified the protocol by using 30uL sample added to 190 uL working solution and a finer-scale linear standard curve (0 - 1.0 ng/uL DNA), which produced good results, as verified by measuring known concentrations of the control Oli molecules. The agreement between this measurement, which is based on the total amount of DNA present, and the initial number of molecules calculated by qPCR were used to calculate the proportion of the initial population that was able to fold and self-prime.

Population-level fitness (increase in population size per PCR cycle) was calculated manually from the linear section of the qPCR curve, and standardized to the rate of increase in population size per PCR cycle of the standards on that plate.

*Sequencing and bioinformatics.* Populations at time points 1, 3, 7 and 11 were pooled by treatment. Inadvertently, replicate populations were not tagged, so information that was expected to be available was lost at this step. While unfortunate for the analysis of this particular experiment, it does not affect the proof of principle that the Oli system and "fold or die" selection may be used for this type of analysis. Because the data shown are from pooled samples, we limit our interpretation of the patterns of diversification and extinction patterns shown. They are meant only to illustrate what the system can do. In principle, individual populations should be tagged, allowing for proper reconstruction of networks below.

Sequencing and bioinformatics were carried out at The GenePool Genomics Facility at the University of Edinburgh. Scripts used are available on request. The pipeline used was:

**1.** Roche 454 Pyro-sequencing: Sequencing libraries were prepared as per the Roche 454 Titanium library preparation protocols (Roche 2009), using the LMW (Low Molecular Weight) library method. Each of the sixteen samples was sequenced in separate lanes of 1/16 or 1/8 of a picotiter plate on the Roche 454-Flx Titanium sequencing platform, using in total just over three plates. Signal processing and base-calling were performed using the Roche shotgun signal-processing software, gsRunProcessor version 2.5.3. Run quality was checked and statistics for the 454 reads were calculated using custom Perl scripts. The first sample (SC02) had been sequenced and reads randomly sampling (eg. at 10000, 20000, 30000 reads) to see at what read numbers new novel sequences stop appearing, in order to get an estimate the number reads required to maintain the same numbers of jMOTU clusters, then the remaining fifteen samples were sequenced to obtain at least this number of reads.

**2.** Extraction of reads: The sff files of reads were converted to fasta file format using the Roche's 'sffinfo' program, with the '-notrim' option to maximise the read length by *not* quality trimming. The number of 454 reads generated for each sample is in Supplementary Table S2.

**3.** Clustering of Reads: Clustering of reads was performed using two different methods of comparison: (1) jMOTU, and (2) a Perl script. First the jMOTU program (Jones, 2011) was used to cluster reads within each sample for sequence differences of 1 to 30 mismatches. The clustering parameters were: "No minimum length", "Low BLAST identity filter" set to 97%, "Percentage of minimum sequence length" set to 87. To run jMOTU, eight gigabytes of memory were allocated to the Java virtual machine using "java -Xmx8000m". jMOTU first uses megaBlast (NCBI, ) to identify pairs of reads with high similarity, then carries out a Needleman-Wunsch alignment to calculate the exact distances between pairs of reads. "MOTU" is an acronym for molecular operational taxonomic units. The second method used custom Perl scripts to: filter reads allowing limited mismatches in the start and end primers; identify all unique sequences by collecting together reads that exactly match including checking the reverse complement of the sequence; then ordering the sequences by the most abundant within each sample and across all the samples. The final number of unique sequences for each sample is listed in the right column of Table S2.

**4.** Restriction-site digestion: In the experiment the sequences were restriction digested before folding, using two enzymes (same restriction site): *Bsa1* which cuts at: 5'GGTCTC(N)3' and 3'CCAGAG(NNNNN)5'; and *HpyAV* which cuts at: 5'CCTTC(NNNNNN)3' and 3'GGAAG(NNNNN)5'. The 454 reads were similarly cut using a Perl script with regular-expressions to trim the reads at these sites.

**5.** Folding prediction: The UNAFold software (Markham and Zuker 2008) was used to predict the secondary structure of the folded nucleic acid sequence for the twenty most abundant sequences within each sample. UNAFold replaces the earlier mfold software, and includes DINAMelt. For its predictions, UNAFold combines free energy minimization, partition function calculations and stochastic sampling. The predicted structures were plotted for comparison. The parameters used for UNAFold were:

       UNAFold.pl --NA=DNA   --temp=55 --max=200   sequenceFile

*Sequence alignments and network analysis.* Sequence alignments and network analysis were carried out on the sequences from the pooled samples above. For each treatment, each unique sequence at the most recent generation was pairwise aligned to each in the previous generation using the Smith-Waterman alignment algorithm (gap opening penalty of 20, extension penalty of 1). The code is available on request. The closest match (with ties broken at random) was recorded for each sequence as its putative ancestor. This was then repeated for each previous generation with the exception of the first. To construct a graphical representation of the population, each unique sequence was placed as vertex with edges provided by the inferred ancestor descendent relationships. Vertices were annotated with generation number and sequence count and edges with the hamming distances. These graphs were visualized using Cytoscape (Shannon et al. 2003). Here, predictors of lineage success were analysed using an ANOVA, where the total number of descendents was used as a measure of lineage success.

**Results**

The results presented here show the range of data types that can be collected from a "fold or die" experiment with Oli populations. Most importantly, Oli populations adapt in response to selection for replication at increasing temperatures. This is shown in Figure 2A. Adaptation is similar to results seen in standard microbial selection experiments, and can be analysed in the same ways. Here, the population-level fitness at any time point is the rate at which the population increases, which is equal to the slope of the qPCR curve during log-linear growth. In all rising temperature treatments, the fitness of the evolved populations is higher than the populations in the control treatment ($F_{3,219} = 9.57$, $p <<0.0001$) (end relative slopes as grand means for 56 populations ± s.d. for each treatment: control = 1.41± 0.0024, one step = 1.54±0.0031, five steps = 1.55±0.0059, ten steps = 1.62±0.0024), with slower rates of environmental change reaching higher end fitnesses ($F_{1,221} = 19.45$, $p<<0.0001$). This is consistent with previous results obtained in digital (Collins et al. 2007) and cellular (Collins and de Meaux 2009) experiments. The most likely phenotypes at 70ºC predicted by mfold of the most successful lineages in each treatment at the beginning of the experiment, along with their most common descendants, are shown in Figure 2B. Since programmes such as mfold ignore three-dimensional structure, and since the folded structure of oligonucleotides can be dynamic, the most likely structure is not necessarily the one used for self-priming during the experiment. Indeed, the structure used for self-priming may be unlikely and/or transient. Here, visualising the most likely structure here under the same conditions demonstrates only that phenotypic evolution has occurred and shows one way in which it can be quantified using either a structural comparisons or differences in folding free energy.

The proportion within populations of individuals that fold and self-prime, as measured by the discrepancy between the total amount of DNA present and the total number of molecules detected by qPCR at each time point, varies over time and between treatments. However, it does not remain high in any of the treatments, despite occasional bouts of wild success in some replicate populations. This can be seen in Figure 2A. Where the among-population variance is high (such as time point 9, in the 10-step treatment), it is because of variation in the proportion of individuals within a population that fold. Since the mutation rate in this system is high, this is not surprising, as any well-adapted types would be unable to maintain adaptation in the face of mutation, which would select for lineages that are robust to mutation. A similar phenomenon is seen in viruses with high mutation rates (Codoner et al. 2006). In subsequent experiments, we recommend using a Taq polymerase with a substantially lower mutation rate to avoid this, unless you wish to carry out a test of selection for "survival of the flattest".

Figure 2

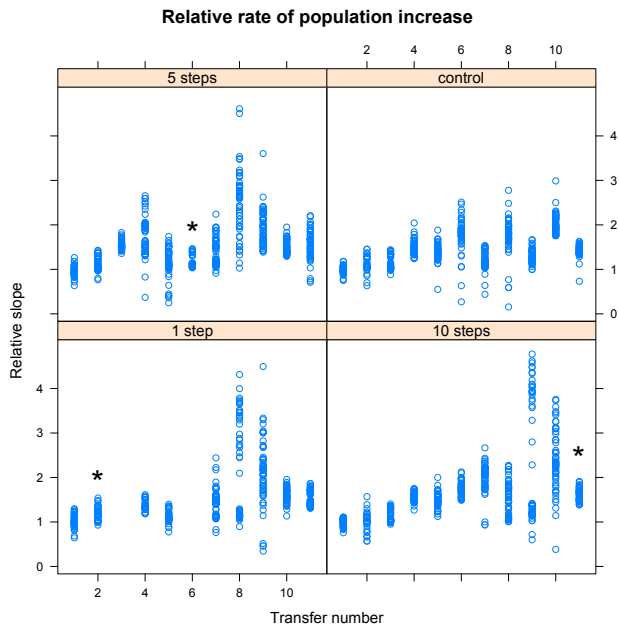
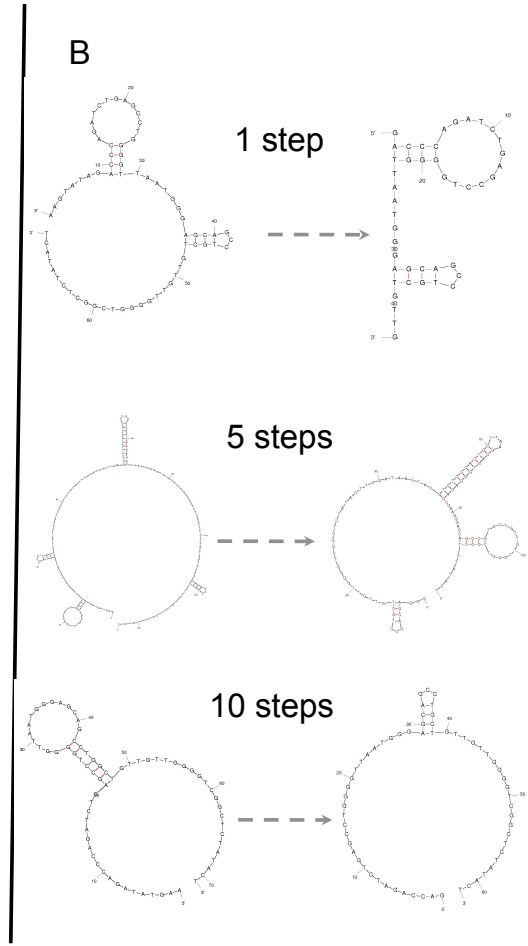

**Figure 2** Fitness trajectories and phenotypic evolution. A: Population growth rates, shown as relative slope, at each cycle of "fold or die" selection. Each dot represents an independent replicate population. The transfer number where the populations reach the maximum temperature of 70ºC is indicated with an asterisk. For temperatures corresponding to individual timepoints in individual treatments, see Table 1. B: Phenotypic evolution. The most likely structure at 70ºC of the most numerous evolved Oli sequence at the end of the experiment (right hand structure) and it's ancestor from transfer 1 (left hand structure), for each treatment. The end of the experiment corresponds to transfer 11 in panel A.

Figure 3

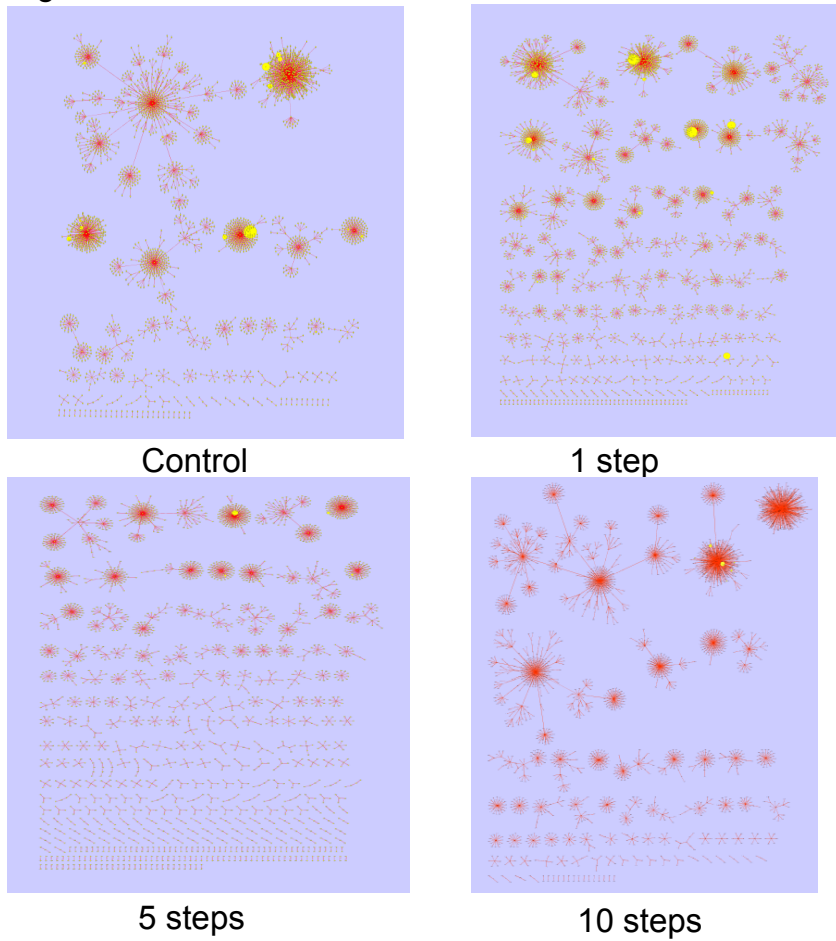

**Figure 3** Network diagrams showing radiations and extinctions in each treatment. Diagrams were made using transfers 1, 3, 7 and 11. Nodes are colour coded by time so that transfer 1= red, transfer 3=yellow, transfer 7=green and transfer 11=violet. Sizes of nodes are proportional to the number of individuals of a particular unique sequence in the pooled populations. Distances between nodes are proportional to differences in sequence. Since networks were made using populations pooled by treatment, they show average dynamics and illustrate one type of data that the Oli system can produce. See the Methods section for a detailed description of how network diagrams were produced.

Figure 3 shows network diagrams depicting radiations and extinctions over time in the different treatments. Lineage success is measured by the number of descendants that a lineage has produced by the end of the experiment. Lineages at time point 1 are referred to as "founders". Note that these founders are one round of selection and amplification (Figure 1) removed from the original Oli sequence, and are all derived from the same Oli ancestor shown in Figure 1.

For illustration, these alignments can be used to understand predictors of lineage persistence. Recall that the particular results and statistics here are for pooled populations. There is no effect of treatment on lineage success ($F_{3,3} = 0.75$, p=0.52) Here, founders with the most descendants are also the ones with the most diverse descendants (highest number of unique sequences at the end of the experiment) ($F_{1,1} = 68.37$, $p < 0.0001$), and this effect varied by treatment ($F_{3,3}=13.99$, $p < 0.0001$). In addition, the number of individuals present in a founding lineage does not predict the eventual success of that lineage in any treatment ($F_{1,1}= 0.029$, p=0.86). Taken together, this suggests that the best strategy for persistence while adapting with a high mutation rate is to produce diverse offspring, some of which will survive both the high mutation rate and the environmental degradation, rather than to produce many similar offspring who then try to crowd out other lineages.

Conversely, extinction rates can be calculated, as shown in Table 1. By the end of the experiment, the control treatment has a population that derives from fewer ancestors than do the high temperature treatments. This appears to be due to bursts of diversity early on in adaptation in the face of environmental change, after which the per-lineage extinction rate is about the same at any time point regardless of treatment. While the treatments have different absolute numbers of lineages, the extinction rate per lineage is the same.

## Summary of extinction events

| Treatment | Time | Number of ancestors | Number of extinctions | Cumulative extinctions | Scaled cumulative extinctions |
|---|---|---|---|---|---|
| control | 1 | 257 | 0 | 0 | 0 |
| control | 2 | 175 | 82 | 82 | 0.319066148 |
| control | 3 | 27 | 148 | 230 | 0.894941634 |
| control | 4 | 8 | 19 | 249 | 0.968871595 |
| 1 step | 1 | 1074 | 0 | 0 | 0 |
| 1 step | 2 | 888 | 186 | 186 | 0.173184358 |
| 1 step | 3 | 106 | 782 | 968 | 0.901303538 |
| 1 step | 4 | 45 | 61 | 1029 | 0.958100559 |
| 5 steps | 1 | 1676 | 0 | 0 | 0 |
| 5 steps | 2 | 1305 | 371 | 371 | 0.221360382 |
| 5 steps | 3 | 113 | 1194 | 1565 | 0.933770883 |
| 5 steps | 4 | 36 | 77 | 1642 | 0.979713604 |
| 10 steps | 1 | 667 | 0 | 0 | 0 |
| 10 steps | 2 | 473 | 194 | 194 | 0.290854573 |
| 10 steps | 3 | 29 | 444 | 638 | 0.956521739 |
| 10 steps | 4 | 16 | 13 | 651 | 0.976011994 |

**Table 1** Summary of extinction events pooled by selection regime. Time refers to the transfer number. The number of ancestors is the total umber of unique sequences from time 1 that are still present within a given selection regime at the time stated. The number of extinctions is the number of unique sequences that have gone extinct between a timpoint and the timpoint immediately preceeding it. Cumulative extinctions is the total number of unique sequences that have gone extinct between time 1 and the timepoint of interest. Scaled cumulative extinctions reflect the total proportion of unique sequences present at timepoint 1 that have gone extinct by the timepoint of interest. No errors are reported since the entire populations, minus the amount used for propagation and measurements during the selection experiment, were surveyed.

Figure 4

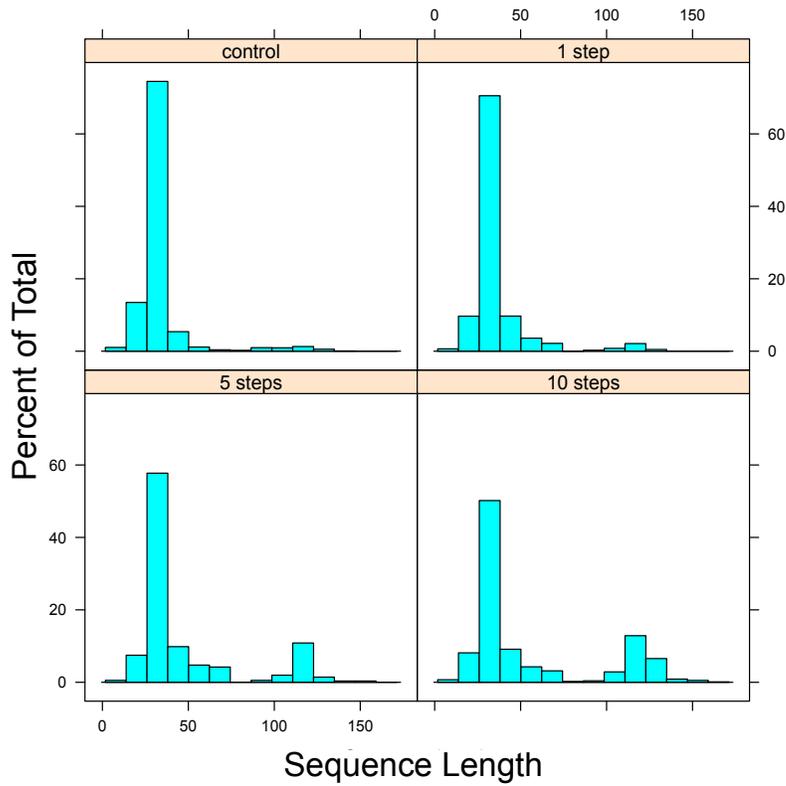

**Figure 4** Distributions of genome lengths in evolved Oli populations, pooled by treatment. Each panel corresponds to one treatment c = control, 1 = one step, 5 = five steps, 10 = ten steps. For reference, the length of the ancestral Oli genome is 107 base pairs long.

One of the striking outcomes of this experiment was the repeated occurrence of drastic changes in genome length. This is summarized in Figure 4**.** Changes in genome size resulted in an important hurdle to producing the networks discussed above, which was aligning palindromic sequences with many insertions/deletions. This could be eased in future experiments by using Taq with a lower error rate and subjecting populations to fewer PCR cycles in between rounds of selection for self-priming. Providing fewer limiting primers and sequencing populations at more tightly spaced time points could further help analyses by providing more intermediate sequences.  The evolution of short sequences is expected in selection experiments using short oligonucleotides, and the "tyranny of short motifs" has been observed before (Ellington 2009).Interestingly, though the end populations in our experiment are dominated by sequences shorter than the ancestor, longer sequences are common in the subset of sequences at intermediate time points that leave descendants through to the end of the experiment. For example, for each treatment, the most common sequence in the final evolved populations has ancestors where the length at time one is 175 base pairs (control), 70 base pairs (one step), 185 base pairs (five steps) and 181 base pairs (ten steps). The same pattern continues when these successful lineages are tracked from time point 1 through to the end of the experiment. This may be because, although short sequences replicate faster during the qPCR reaction, they may also be at a disadvantage during the fold or die step of selection.  In addition, different selection regimes result in different proportions of long and short molecules evolving, where populations evolved under gradual rates of environmental change have a higher proportion of long molecules than either the control populations, or those evolved under sudden environmental change (proportion of sequences over 75 base pairs in end populations for each treatment: control = 0.042, 1 step = 0.038, 5 steps = 0.183, 10 steps = 0.324). This suggests that the extent of the "tyranny of short motifs" that emerges during experiments using oligonucleotides depends on the strength of selection present.

**Discussion**

The main result presented here is that the Oli system and "fold or die" selection can be used for experimental evolution in much the same way as viral or bacterial populations, using resources and expertise available in most microbial evolution labs. The system has both advantages and limitations relative to living systems. The main advantages of this system are that the populations are completely tractable genetically, phenotypically, and in terms of fitness. Moreover, any individual sequence, including rare or extinct ones, or even ones that did not occur, can be easily synthesized and its fitness measured. The ability to create individuals that never occurred during highly-replicated adaptation events allows us to directly test hypotheses about the presence and nature of evolutionary constraints, while the possibility of re-creating intermediate types in any desired order allows us to directly measure the fitness distributions consequences sign epistasis between sequential mutations. Distributions of fitness effects (mutation, epistatic, pleiotropic, genome size) can be determined empirically rather than estimated from a subsample that excludes rare types or events. This means that the Oli system can produce data that will improve the distributions of fitness effects that we use to inform our fundamental ideas about the way adaptation works, including the importance of rare events. Though not done here, Oli populations could also be used to investigate the emergence of cooperation (using others to prime and extend) or to explore questions about the evolution of recombination rates. Here, we used populations that were initially dilute to minimize the advantages of cooperation and recombination, but there is no reason that others could not change this. When Oli is used in conjunction with other (live) systems, these many advantages outweigh the limits and faults of the system, and these advantages should increase as improvements to the system are made.

All experimental systems have limitations, and Oli is no different. Most obviously, Oli is not alive, and lacks the complexity of even the simplest living system. However, the point of experimental evolution is rarely to reproduce natural systems in all their glorious complexity, but rather to simplify them enough so that we can understand the general rules that evolution plays by. A second limitation is that we have not yet been able to repress cooperation and recombination in this system (nor has it been quantified it in this experiment – this is not an inherent limit of the system but rather a decision about what to include in the first experiment with it). As such, Oli populations are not, and at present cannot be, completely asexual. One cheeky way around this problem would be to turn it on its head and simply use Oli populations to study the evolution of cooperation and the evolution of recombination rates. Finally, Oli populations are tricky to sequence. Since fold-or-die selection results in ever more stable hairpins, and because the precise outcome of evolution is unknown, sequencing protocols must be modified and tested at the end of the experiment. This requires a good working relationship with a sequencing centre. However, since Oli hairpins must melt so that individuals can reproduce in the PCR reactions, the hairpins are unlikely to evolve stability that makes sequencing impossible.

We expect to make, and hope that others will make, improvements to the Oli system. The first and most pressing improvement would be to use a polymerase with a lower mutation rate, so that more intermediate genomes are available during adaptive walks. However, polymerases with lower mutation rates do not have even mutation spectra, which introduces a mutational bias into the experiment. This could be accounted for at the end. A second and more ambitious improvement would be to estimate and eventually control the recombination and concatamerization rates of Oli molecules in the PCR reactions. While recombination seems unlikely early in the PCR program when populations are dilute, it becomes increasingly likely as population density increases. Since Oli sequences must be partially palindromic and self-similar to survive fold-or-die selection, recombination becomes likely, as does the use of other Oli molecules as primers. This results in concatamerization and changes in genome size by duplicating or deleting entire segments, and may also introduce an element of genome "modularity" into the system. A final and powerful improvement would be the design of several different starting Oli sequences to allow for true replication in experiments, which is necessary to generalize results. While most microbial experimental evolution uses a single genetic starting point and still produces generalizable results, the outcome of experiments is always at least partially contingent on the genetic starting point, especially when selection is weak (Travisano et al. 1995; Collins et al. 2006). One aspect of the Oli system that remains to be tested is how well the realized structure of the folded molecules matches the structures predicted by folding programs. Oli is based on a structure where hairpin formation was empirically tested, and must be able to at least function as if they can fold, but it is unknown how well the most likely structure predicted by folding programs predicts the actual phenotypes present in this system.

The idea of using short oligonucleotides to investigate evolutionary phenomena is not new. It is usually applied to the evolution of hypercycles to address variations on origin of life questions laid out by Maynard-Smith (Smith 1979) (But see (Wlotzka and McCaskill 1997) for an interesting investigation of preditor-prey systems using self-sustained sequence replication). In contrast, the Oli system provides a novel tool for investigating evolution at the genetic, genomic, phenotypic and fitness level in populations under conditions that are simpler and better-defined than those needed for viruses or cellular organisms. Right now, accessible high-throughput sequencing is giving us more data than ever before on the genetic and genomic changes that occur during evolution, but it doesn't let us map those changes to the changes in fitness that natural selection ultimately acts on, or to see how natural selection triages variation that is produced. The Oli system provides a way to construct genotype-phenotype-fitness maps under a range of selection and demographic regimes. One apt comment on the growing use of new technologies such as high-throughput sequencing is that "The techniques have galloped ahead of the concepts. We have moved away from studying the complexity of the organism; from processes and organization to composition." (in Andrew Jack, "An Acute Talent for Innovation", Financial Times (1Feb 2009)). The new "normal" problem in evolutionary biology is a giant mountain of sequence data, from which biologists are supposed to extract meaningful insights about the world outside our computers and labs. The Oli system uses the tools that are burying us in data about composition, and subverts them to learn about process, which is not only useful, but also very satisfying indeed.

**Acknowledgements**
We thank Heidi Kuehne for laboratory assistance; Petra Schneider and Andrea Betancourt for discussions on how best to subvert qPCR; Mark Blaxter, Karim Gharbi, Anna Montazam, and Denis Cleven at The GenePool Genomics Facility for experimental design, library preparation and sequencing, and Martin Jones for advice on jMOTU. This work was supported by NERC small project grant NE/G00904X/1 to SC.

# Supplementary Figures

## Self-priming temperatures

| Transfer number | Selection regime | | | |
|---|---|---|---|---|
| | Control | 1 step | 5 steps | 10 steps |
| 1 | 55 | 55 | 55 | 55 |
| 2 | 55 | **70** | **58** | **56.5** |
| 3 | 55 | 70 | 61 | 58 |
| 4 | 55 | 70 | **64** | **59.5** |
| 5 | 55 | 70 | **67** | **61** |
| 6 | 55 | 70 | **70** | **62.5** |
| 7 | 55 | 70 | 70 | **64** |
| 8 | 55 | 70 | 70 | **65.5** |
| 9 | 55 | 70 | 70 | **67** |
| 10 | 55 | 70 | 70 | **68.5** |
| 11 | 55 | 70 | 70 | **70** |

**Table S1** Temperatures in °C used for the self-priming step at each transfer for each selection regime. The total magnitude of change in temperature is the same for all selection regimes over the course of the experiments. Only the rate of change varies.

## Number of 454 reads generated for each sample

| Population | Sample name | Number of reads | Mean read length | Number of unique sequences |
|---|---|---|---|---|
| control(t1) | SC02 | 94137 | 138.9 | 27420 |
| 1 step(t1) | SC57 | 30077 | 157.6 | 12040 |
| 5 steps(t1) | SC58 | 38979 | 161.1 | 16880 |
| 10 steps(t1) | SC59 | 63670 | 142.5 | 27961 |
| control(t3) | SC60 | 27808 | 158.6 | 11493 |
| 1 step(t3) | SC61 | 24893 | 156.3 | 10690 |
| 5 steps(t3) | SC62 | 29103 | 156.6 | 13611 |
| 10 steps(t3) | SC63 | 45018 | 160.3 | 19831 |
| control(t7) | SC64 | 22648 | 134 | 15033 |
| 1 step(t7) | SC65 | 34025 | 141.6 | 18533 |
| 5 steps(t7) | SC66 | 41454 | 131.5 | 21999 |
| 10 steps(t7) | SC67 | 38322 | 124.4 | 18616 |
| control(t11) | SC68 | 67095 | 117.4 | 20495 |
| 1 step(t11) | SC69 | 79143 | 117.7 | 25131 |
| 5 steps(t11) | SC70 | 21007 | 121.9 | 11379 |
| 10 steps(t11) | SC71 | 24211 | 122.7 | 13694 |

**Table S2** Number of 454 reads generated for each sample. Each sample refers to pooled populations at a single timepoint. Populations are referred to as selection treatment(transfer number). For example, the pooled control populations at time 1 are denoted control(t1), the populations from the 1 step treatment at time 1 are denoted 1 step(t1) etc. Sample name is the short name of the sample used in the data archive.